\documentclass[11pt]{article}

\usepackage{times}
\usepackage{eurosym}
\usepackage{amssymb}
\usepackage{amsmath}
\usepackage{latexsym}
\usepackage{stmaryrd}
\usepackage{soul}
\usepackage{graphicx}
\usepackage{color}
\usepackage{natbib}
\usepackage{pifont}

\newtheorem{definition}{Definition}[section]
\newtheorem{proposition}[definition]{Proposition}
\newtheorem{lemma}[definition]{Lemma}
\newtheorem{theorem}[definition]{Theorem}
\newtheorem{remark}[definition]{Remark}
\newtheorem{corollary}[definition]{Corollary}

\newcommand{\NN}{\mathbb{N}}

\newcommand{\calA}{\mathcal{A}}

\newcommand{\sort}[2]{\mathcal{S}^{#1}_{#2}}
\newcommand{\filter}[1]{\mathcal{F}_{#1}}

\newcommand{\first}{{\rm first}}

\newcommand{\sem}[1]{[\![#1]\!]}

\newcommand{\price}{{\rm price}}
\newcommand{\rating}{{\rm rating}}
\newcommand{\size}{{\rm size}}
\newcommand{\noRooms}{{\rm rooms}}
\newcommand{\dimension}{{\rm dimension}}
\newcommand{\capacity}{{\rm capacity}}
\newcommand{\threebedor}{{\rm three\_beds\_or\_near\_station}}
\newcommand{\bedrooms}{{\rm bedrooms}}
\newcommand{\distance}{{\rm distance}}

\newcommand{\List}[1]{{\bf List} \, #1}

\begin{document} 

\title{Sorting and filtering as effective rational choice procedures}

\author{Paulo Oliva and Philipp Zahn}

\maketitle

\begin{abstract}
  Many online shops offer functionality that help their customers navigate the
  available alternatives. For instance, options to filter and to sort goods are wide-spread. In this paper we show that sorting and filtering can be used by rational consumers to find their most
  preferred choice -- quickly.   We characterize the
  preferences which can be expressed through filtering and sorting and show that
  these preferences exhibit a simple and intuitive logical structure.

\end{abstract}	


\section{Introduction}

Many online shops offer functionality for consumers to find their
preferred alternative. Filter and sorting in particular are
ubiquitous among e-commerce shops. 
In this paper we ask: how does  filtering and sorting support the search process of a
rational consumer, i.e. a consumer who has a preference relation and optimizes
accordingly.

At first, this may sound contradictory. How can such systems help rational
agents? After all, a rational agent has to inspect
all the alternatives in order to ensure to choose the optimal one. Yet, as we
show, a combination of filtering and sorting and then choosing the first element of the resulting list
implements a rationalizable choice. 

Furthermore, preferences which can be expressed through filtering and sorting
have a simple logical structure and can be described in a
natural way. As an example, an agent expressing his preferences for an external
hard drive could state the following: ``I prefer hard drives with bigger capacity
\emph{and} prefer the Seagate brand over other brands''.  
Importantly, these preferences bear
resemblance with non-compensatory,
conjunctive decision rules discussed in the marketing literature (e.g.
\cite{gilbride2004choice}).\footnote{The logical structure of such preferences
  is also reminiscent of the discussion
  in \cite{Rubinstein1998definable}.} 

Of course, it is not clear at all why individuals should hold such preferences. Hence,
we consider two extensions. First, if preferences are defined on attributes but a particular attribute is
not displayed, a combination of filtering, sorting and
satisficing according to the missing attributes leads to the optimal choice.
 Secondly, we show how extending
the filter functionality, by adding a union operation (``look for houses with
either 5 bedrooms \emph{or} located in neighborhood x''), will substantially
increase the expressiveness of the preferences that our environment allows for.
Doing so pushes more complexity away from the agent into the functionality
supporting the choice process. Analogously to the conjunctive decision rule, the
union operation and the type of preferences it induces do have a counterpart in
decision rules: disjunctive rules \citep{gilbride2004choice} and, as we can
combine these structures, also  ``disjunctions of conjunctions'' \citep{hauser2010disjunctions}.   

The bottom line is: Sorting and filtering allow an agent to choose more
effectively among the alternatives. It quickens the choice process
by allowing the agent to ignore large chunks of the available alternatives.
Comparing it to a benchmark where the agent fully optimizes, we
derive a simple formula,  which indicates when 
using the sorting and filtering is ``quicker'' than inspecting all elements.  The fewer attributes the agent cares about and the larger the number
of alternatives, the more effective the functionality is. 

The implications for an online shop are immediate. In so far as it can provide
adequate decision support, it can offer more alternatives without increasing the
complexity of finding the optimal alternative. And in so far as consumers like variety, having adequate support 
thus allows to increase variety (and may induce a competitive advantage over
other platforms).

In our model we intentionally focus on a rational consumer as a benchmark. This
allows us to exclude any other channel through which decision support may play a role. In
particular, we ignore fundamental rationality limitations or biases on the
agent's side. But it is clear that a well designed choice support
also helps consumers who are not perfectly rational to make
decisions. Consider an agent overwhelmed by the number
of choices, i.e. experiencing choice overload \cite{iyengar2000choice}. An effective decision support system may help such a consumer
to decide. 

In our model, we focus on the simple functionality of sorting and filtering, as
they are widely employed. Yet, the techniques we use to characterize decision
support and its relation to preferences could be applied to other settings as
well. What is required is that the operations that can be performed are limited.
In the context of online shops this, of course, is given by design. The choice
operations in essence define a simple ``choice
language grammar'' whose semantics can be analyzed using standard tools from theoretical
computer science and can then be related to preference relations (or other modes
of decision-making).


\paragraph{Related literature}

From a bird's eye perspective, one can interpret our abstract treatment of filtering and sorting as
instances of a ``choice environment'' which influences the choice process. This
idea goes back to Herbert Simon \citep{Simon1956}. He argued that agents do not make decisions in the void but are
embedded in an environment which contains information and which affects
their choice procedure \cite{Simon1956}. In particular, if the
environment is sufficiently rich in relevant information, procedures can be
simple but still achieve complex goals. While Simon's ideas on \emph{internal}
decision-making constraints of agents \cite{Simon1955} has received a lot of attention in the
literature on bounded rationality, to the best of our knowledge, the choice environment has been
mostly ignored in that literature. One exception is \cite{Todd2003}: They
also recognize the lack of considering the environment in the choice process. While sharing this point, our paper otherwise has little
in common. We focus on a formal treatment and rational choice for a limited
environment to illustrate the importance of considering the environment; they use an
informal treatment and focus on simple heuristics over a broad set of different environments.

Our paper is related to several themes in the marketing literature. First,
empirical evidence indicates that product attributes are
relevant in the search process. \cite{bronnenberg2016zooming} document search patterns for
customers looking for a camera online.  They find that the attributes customers
search for are highly informative for the ultimate buy decisions. 
Moreover, it supports the view that steps in the decision process
which involve product attributes can
be used to deduce preferences.\footnote{In fact, in the presence of search costs \emph{not} taking into account the
search process may lead to biased evaluations of consumer choice
(cf. \cite{koulayev2014search}).}

That attributes matter in the search process is also supported by studies on
search queries and ensuing choice behavior.  \cite{liu2018semantic} investigate how
search expressions match agents' preferences. They find that search content is
highly informative of preferences.  \cite{shi2021path} provide a detailed documentation of the
information-processing that agents perform when choosing on results. Agents are
affected by several variables including visual context (e.g. spatial orientation
on the page). Most relevant for us, agents are
influenced by the semantic content of the search results in particular if
they are searching ``transactionally''. It means, they are looking for a
specific object or service to buy -- like in our setting.  In sum, attribute content is key when agents navigate through a platform.

A second strand of the literature relevant for our paper concerns the choice
environment. In the presence of search costs effective decision support helps
customers making better decisions. Empirical estimates of search costs for single-attribute search (\cite{hortaccsu2004product,de2012testing,santos2017search}) but also
for multi-attribute search (\cite{moraga2013semi}) suggest that customers
exhibit significant search costs.\footnote{The availability of detailed data on
  online platforms is also leveraged to empirically
  evaluate theoretical search models. See, for instance,
  \cite{kim2017probit} and \cite{chen2017sequential}.} They also typically engage in
limited search. Hence, the easier it is for agents to navigate an online shop,
the quicker they will be able to find their preferred alternative thereby
reducing search costs. Focusing on two types of decision aids,
\cite{haubl2000consumer} document how such aids help consumers make better
(non-dominated) choices while at the same time lowering 
 search costs. How search costs can be reduced is investigated in \cite{de2008offering} who devise a preference
elicitation questionnaire based on conjoint analysis including background
information on the agent's side. Similar to the ideas in
our paper, the goal is to reduce the number of required steps a customer needs
to make in order find his/her optimal product. In contrast though, our paper is focused
on the attributes of the products as well as functionality widely available on
webshops and is not investigating how customer
background information can be leveraged.

Through the focus on rational preference relations, our paper also touches on
themes in economic decision theory. First, our paper is related to a recent literature on
search behavior, for  instance \cite{caplin2011search} and \cite{masatlioglu2013choice}. These 
papers investigate the search process of an agent and how it relates to his
underlying preferences. Our approach differs as the environment is not just
providing passive information, such as an order of alternatives or a recommended
element based on some statistic, but an agent can make use of functionality
provided through the environment.  Second, our paper is also related to two papers by Ariel Rubinstein and Yuval Salant
\cite{Rubinstein2006b,Salant2008}. In both papers, they consider decision
problems with extra structure that can be used by an agent's choice procedure
and then ask under which conditions such a procedure can be rationalized. In
\cite{Rubinstein2006b} they consider the case where alternatives are
organized as lists; in \cite{Salant2008}  they consider additional information captured in
``frames'' which affects the decision-maker's choice. Possible frames include
the representation of alternatives in the form of a list. In our model,
alternatives are also organized in lists and this also plays a key role in
determining choices. While in their case operations on lists are internal, we
consider functionality provided from outside in addition to the
representation of alternatives in lists. 
Lastly, this paper is  building on results developed in \cite{OlivaZahn2018} where a general model of the environment which contains information regarding
alternatives is introduced. Using abstract representations, the authors characterize which representations and choice
procedures can be rationalized. There are three differences to our paper. First,
the environment model is passive and there is no
functionality an agent can employ. Second, the results are not constructive, i.e. it is shown that a rational
preference relation exists without characterizing it. Third, the complexity of
decision procedures is not considered. Lastly, in our paper we keep the
technical necessities at a minimum, for instance, we only consider one
representation of alternatives in the form of lists.    

\section{Basic Setup and Decision Support}

Let us first informally state the scenario  which we are interested in and then
turn to the nuts and bolts. Picture a customer who wants to buy a certain type of
good or service $x$ on an online platform. He has rational preferences and, after searching
a platform, such as Amazon or AirBnB, faces a list of results. The platform
gives him a list of $n$ results and offers him decision support by
showing attributes according to which he can filter the results as well as criteria
to sort the results.   

Our first task is to describe the choice environment and the functionality offered by it. 

\subsection{Alternatives and attributes}

\begin{definition}[Attribute] Let $X$ denote the overall set of alternatives and $(V, \succeq)$ be a totally ordered set. We call a function $a \colon X \to V$ an \emph{attribute map} or simply an \emph{attribute}. 
\end{definition}

For instance, if we take $X$ as a set of shoes, and model sizes using natural
numbers with the standard ordering $(\NN, \geq)$, we can think of a mapping
$\size \colon X \to \NN$ as assigning sizes to each pair of shoes in $X$.
An agent may also may consider the quality of goods and rely on ratings, say
defined for the real interval [1,5] (again with the standard ordering). Then, we
can also think of the mapping $\rating \colon X \to [1,5]$ as the average review
$a(x) \in [1,5]$ of each pair of shoes $x \in X$, assuming customers will rate
the quality of shoes using values $\{1,2,3,4,5\}$. 

\begin{remark}
In the example above, attributes describe simple properties. We call such
attributes \emph{atomic}. We note here, attributes can be more complex. Consider
a family looking for a holiday home. They look for a quite place, with garden
and pool, and easily accessible by car. A platform may offer a composite
\emph{attribute}, ``family friendly'', which combines existing attributes in a
particular way. 
\end{remark}

\subsection{Filtering and sorting}

In this paper the decision problem is given to the agent not as a set, but as a list of alternatives. If $X$ is the overall set of alternatives, we write $\List{X}$ for the set of finite lists over $X$.

\begin{definition}[Filtering] Given any attribute map $a \colon X \to V$ and a value $v \in V$, the \emph{$a$-higher-than-$v$-filter} function $\filter{a \geq v} \colon \List{X} \to \List{X}$ is the function which filters the elements of a list according to the predicate $a(x) \geq v$. The output list $l' = \filter{a \geq v}(l)$ should satisfy the following properties:
\begin{itemize}
	\item[$(F_1)$] $l'$ consists precisely of the elements $x$ of $l$ which satisfy $a(x) \geq v$.
	\item[$(F_2)$] if two elements $l_i$ and $l_j$ satisfy the filtering property and $i < j$ then the order of these two elements is preserved in $l'$.
\end{itemize}
The two conditions together ensure this uniquely defines the function $\filter{a \geq v}$. Dually, we can also define the \emph{$a$-lower-than-$v$-filter} function $\filter{a \leq v} \colon \List{X} \to \List{X}$ satisfying similar properties.
\end{definition}

So, for instance, we can talk about the $\size$-lower-than-10-filter
$\filter{\size \leq10}$. It selects shoes which are size 10 or smaller.
Or we can talk about the $\noRooms$-higher-than-3-filter $\filter{\noRooms \geq
  3}$, which selects houses with at least three rooms.

\begin{definition}[Sorting] Given any attribute map $a \colon X \to V$ we define the \emph{increasing ordering based on $a$} (also called the \emph{increasing-$a$-ordering}) as the mapping $\sort{i}{a} \colon \List{X} \to \List{X}$ which sorts the elements of a list in increasing order according to the value $a(x) \in V$. The list $l' = \sort{i}{a}(l)$ should satisfy the following three properties:
\begin{itemize}
	\item[$(S_1)$] $l'$ be a permutation of the list $l$,
	\item[$(S_2)$] for $i < j$ we must have $a(l'_i) \leq a(l'_j)$, and
	\item[$(S_3)$] if two elements $l_i$ and $l_j$ are such that $a(l_i) = a(l_j)$ and $i < j$ then the order of these two elements is preserved in $l'$.
\end{itemize}
These three properties uniquely define $\sort{i}{a}$. Analogously we can also talk about the \emph{decreasing ordering based on $a$} (or the \emph{decreasing-$a$-ordering}), i.e. $\sort{d}{a} \colon \List{X} \to \List{X}$.
\end{definition}

For instance, we can talk about the increasing-$\price$-ordering or the decreasing-$\rating$-ordering maps. Our first goal is to study the composition of these sort and filter operations. 

\begin{remark} For this paper we are assuming that the value sets $(V, \geq)$
  are totally ordered sets, so that both the filtering and the sorting operations are well defined.
  Although filtering would technically only required $(V, \geq)$ to be a partial order, it is simpler to
  assume that both filtering and sorting make use of the same ordered value set. 
  We note that there are cases where no such (natural) total ordering exists. For instance, suppose
  $V$ is a set of \emph{colors}, e.g. $V = \{$ blue, red, yellow $\}$.
  In these cases one can still treat these within our framework by imposing any fixed total ordering
  on these, e.g. lexicographical ordering. 
  This would allow an agent to filter a particular color by composing $\filter{a
    \leq v}$ and $\filter{a \geq v}$. There are, of course, other ways to deal
  with this technicality but they would make the setup more complex without
  adding much value.
\end{remark}

\subsection{Decision procedures and equivalent choices}

Equipped with filtering and sorting, we can define a decision procedure as a
sequence of these operations. We then show that for each such procedure there
exists a canonical and ``minimal'' form. This, in turn, allows to identify the
most effective way of choosing given these operations.

\begin{definition}[Decision procedures] We will refer to the composition of
  different filtering and sorting operations as a \emph{decision procedure}. The
  number of filter and sort operations used will be called the \emph{length} of the procedure.
\end{definition}

Although sorting and filtering alone will not produce a single choice at the
end, but a (possibly smaller) ordered list, they form the core of how people
choose online. For example, an agent first filters houses with sufficiently many
rooms, and then picks the cheapest element. Assuming an
operation which picks the first element of a list, i.e.
\[ \first \colon \List{X} \to X ,\] 
the procedure above can be described by the following composition
\[
\begin{array}{lcl}
\first \circ \sort{i}{\price} \circ \filter{\noRooms \geq 3} & \colon & \List{X} \to X.
\end{array}
\]
This composition of operations selects a cheapest house (not necessarily unique)
which has at least three rooms $3$ or more. When the filtering is too
restrictive, an empty list results. For now, we will not worry about this
problem. We come back to it later.   

Given a fixed set of attributes $\calA$, let us define a notion of
``equivalence" between decision procedures. We will also consider the
equivalence of alternatives and lists -- clearly prerequisites for equivalence of
decision procedures.

\begin{definition}[$\calA$-equivalence] Fix a set of attributes $\calA$:
\begin{itemize}
	\item Two alternatives $x, y \in X$ are \emph{$\calA$-equivalent}, written $x \simeq_{\calA} y$, if $a(x) = a(y)$ for all attributes $a \in \calA$.
	\item Two lists $l, l' \in \List{X}$ are \emph{$\calA$-equivalent}, written $l \simeq_\calA l'$, if they have the same length and their corresponding elements are $\calA$-equivalent. It is easy to see that $\simeq_{\calA}$ is an equivalence relation on $\List{X}$ and we will write $\sem{l}_{\calA}$ for the equivalence class of $l \in \List{X}$.
	\item Two procedures $P_1, P_2 \colon \List{X} \to \List{X}$ are \emph{$\calA$-equivalent}, written $P_1 \simeq_{\calA} P_2$, if for all lists $l \in \List{X}$ we have that $P_1(l) \simeq_\calA P_2(l)$.
\end{itemize}
\end{definition}

We can derive two immediate lemmas. 

\begin{lemma} \label{equiv class lemma} Let $P, P_1, P_2$ be decision procedures defined over the set of attributes $\calA$:
\begin{itemize}
	\item[$(i)$] If $l_1 \simeq_\calA l_2$, then $P(l_1) \simeq_\calA P(l_2)$.
	\item[$(ii)$] If $l_1 \simeq_\calA l_2$ and $P_1 \simeq_{\calA} P_2$ then $P_1(l_1) \simeq_\calA P_2(l_2)$.
\end{itemize}
\end{lemma}
{\bf Proof}. $(i)$ By induction on the structure of $P$. If $P$ is one of the basic operations (filter or sorting) the result is obvious, e.g. clearly $\filter{a \geq v}(l_1) \simeq_\calA \filter{a \geq v}(l_2)$ and $\sort{i}{a}(l_1) \simeq_\calA \sort{i}{a}(l_2)$ when $l_1 \simeq_\calA l_2$. Now, if $P = P_1 \circ P_2$, by induction hypothesis we have $P_1(P_2(l_1)) \simeq_\calA P_1(P_2(l_2))$, given $l_1 \simeq_\calA l_2$. \\[1mm]
$(ii)$ Follows directly from $(i)$ and the definition of $P_1 \simeq_{\calA} P_2$ as $P_1(l_1) \simeq_\calA P_1(l_2) \simeq_\calA P_2(l_2)$. \hfill $\Box$ \\

Given any attribute $a \in \calA$, let us write $x \succeq_a y$ for $a(x) \geq a(y)$, and $x \preceq_a y$ for $a(x) \leq a(y)$. We will use $\phi, \phi_1, \phi_2, \ldots$ as meta-variables ranging over the predicates $a \geq v$ or $a \leq v$. We start by investigating some algebraic properties of composition of decision procedures with respect to this equivalence relation:

\begin{proposition} \label{prop-swap} Let $\calA$ be a given set of attributes, and $a, a_1, a_2 \in \calA$, and $v, v_1, v_2 \in V$. We have that 
\begin{itemize}
	\item [$(i)$] $\filter{\phi_1} \circ \filter{\phi_2} \simeq_{\calA} \filter{\phi_2} \circ \filter{\phi_1}$, for all $\phi_1, \phi_2 \in \{ a_1 \geq v_1, a_2 \geq v_2, a_3 \leq v_3, a_4 \leq v_4 \}$
	\item [$(ii)$] $\filter{a \geq v_1} \circ \filter{a \geq v_2} \simeq_{\calA} \filter{a \geq \max\{v_1, v_2\}}$ and $\filter{a \leq v_1} \circ \filter{a \leq v_2} \simeq_{\calA} \filter{a \leq \min\{v_1, v_2\}}$
	\item [$(iii)$] $\filter{\phi} \circ \sort{\alpha}{a} \simeq_{\calA} \sort{\alpha}{a} \circ \filter{\phi}$, for $\alpha \in \{i,d\}$ and $\phi \in \{ a_1 \geq v, a_1 \leq v \}$.
	\item [$(iv)$] $P_1 \simeq_{\calA} P_2$ and $P_1' \simeq_{\calA} P_2'$ then $P_1 \circ P_1' \simeq_{\calA} P_2 \circ P_2'$, for arbitrary procedures $P_1, P_1', P_2, P_2'$
	\item [$(v)$] $\sort{\alpha}{a} \circ P \circ \sort{\beta}{a} \simeq_{\calA} \sort{\alpha}{a} \circ P$, for $\alpha, \beta \in \{i,d\}$ and $P$ an arbitrary procedure
\end{itemize}
\end{proposition}
{\bf Proof}. $(i)$ Clearly the order in which one applies two filters to a list does not matter. The two resulting lists will actually be identical, and hence $\calA$-equivalent for any attribute set $\calA$. \\[1mm]
$(ii)$ Easy. Note that $\filter{a \geq v_1} \circ \filter{a \leq v_2}$, however, cannot in general be simplified as it is specifying a genuine interval of preference -- except when $v_2 < v_1$ in which case this would always return the empty list.  \\[1mm]
$(iii)$ Consider the case when $\alpha = i$ and $\phi = a_1 \geq v$, the other case is handled similarly. Let $l \in \List{X}$ and assume $l = \filter{a_1 \geq v}(\sort{i}{a}(l))$ and $l' = \sort{i}{a}(\filter{a_1 \geq v}(l))$. Again, it is easy to see that in this case $l = l'$, and in particular they are $\calA$-equivalent. \\[1mm]
$(iv)$ Easy consequence of Lemma \ref{equiv class lemma}. Indeed, given a list $l$, by the assumption $P_1' \simeq_{\calA} P_2'$ we have $P_1'(l) \simeq_\calA P_2'(l)$. By Lemma \ref{equiv class lemma} $(ii)$ and the assumption $P_1 \simeq_{\calA} P_2$ we obtain $P_1(P_1'(l)) \simeq_\calA P_2(P_2'(l))$. \\[1mm]
$(v)$ We are claiming here that sorting with respect to an attribute $a$, and then later sorting with respect to the same attribute (possibly in the opposite direction) makes the initial sorting irrelevant. By $(i)$, $(iii)$ and $(iv)$, we can assume that $P$ only contains sorting operations, i.e. we must show
\[ \sort{\alpha}{a} \circ \sort{\alpha_1}{a_1} \circ \ldots \circ \sort{\alpha_n}{a_n}\circ \sort{\beta}{a} \simeq_{\calA} \sort{\alpha}{a} \circ \sort{\alpha_1}{a_1} \circ \ldots \circ \sort{\alpha_n}{a_n} \]
Moreover, by induction we can assume that all $a_i$ are different from $a$. Now, one can see that the procedure $P \equiv \sort{\alpha_1}{a_1} \circ \ldots \sort{\alpha_n}{a_n}$ is ordering the elements of any given list according to the lexicographical ordering on the relations $R_i \in \{ \preceq_{a_i} , \succeq_{a_i}\}$ depending on whether $\alpha_i$ is $i$ or $d$. In the case $\sort{\alpha}{a} \circ P$ we are sorting by $a$, and using the lexicographical ordering of $P$ for attributes that have the same $a$ value. In the case of $\sort{\alpha}{a} \circ P \circ \sort{\beta}{a}$ one is also first sorting by $a$, and then by the lexicographical ordering of $P$ for elements that have the same $a$ value, and finally sorting by $a$ again in case of tie-break. But this final step is indeed redundant since we will only apply these to elements which already have the same $a$ value. \hfill $\Box$

Equipped with the proposition before, we can now associate with any decision
procedure a canonical equivalent which minimizes the number of steps needed.

\begin{theorem}[Normal form] We say that two filtering operations $\filter{\phi_1}$ and $\filter{\phi_2}$ are \emph{dual} if $\phi_1 = a \geq v_1$ and $\phi_2 = a \leq v_2$. Any decision procedure is $\calA$-equivalent to a procedure where:
\begin{enumerate}
	\item all filtering operations precede all sorting operations,
	\item all sorting operations are done with respect to distinct attributes, and
	\item at most two dual filtering operations are used for each attribute.
\end{enumerate}
\end{theorem}
{\bf Proof}. Straightforward from Proposition \ref{prop-swap}. \hfill $\Box$ \\

As we can derive for each procedure its normal form, we can also derive the number of steps needed to compute a choice. 

\begin{corollary} \label{cor-upper-bound} Let $N = | \calA |$. Any decision procedure is $\calA$-equivalent to a procedure of length at most $3 N$.
\end{corollary}

For instance, the decision procedure
\[ \sort{d}{\rating} \circ \sort{i}{\price} \circ \filter{\noRooms \geq 4} \circ \sort{d}{\price} \]
is $\{\price,\noRooms,\rating\}$-equivalent to
\[ \sort{d}{\rating} \circ \sort{i}{\price} \circ \filter{\noRooms \geq 4} \]
which indicates preference for the cheapest, highest-ranked house with more than
4 rooms. We consider this to be the normal form of the
procedure above, and this normal form procedure has length $3$.

In summary, for all procedures we can derive a canonical normal form. Moreover, note that if we fix an ordering on the set of attributes $\calA$, and put all filtering operations in this order, the normal form of the procedure will be unique. This result will become handy in the next section when we characterize what type of preferences can result from sorting and filtering. 


\section{Rationalizable Preference Relations}
\label{sec:modelling_pref}
So far, we described the environment and decision procedures operating on it. We
also characterized these procedures through its normal form. We want to relate
procedures to the maximization of a rational preference relation. 

\subsection{Preference relations defined on attributes}

We begin by characterizing preferences in our setting. A preference relation
$\succeq$ is a subset of $X \times X$ satisfying the two rationality conditions: \emph{completeness} (either $x \succeq y$ or $y \succeq x$) and \emph{transitivity} ($x \succeq y$ and $y \succeq z$ implies $x \succeq z$). When $x \succeq y$ we say that $x$ is \emph{weakly preferred} over $y$. One is \emph{indifferent} between $x$ and $y$ when $x \succeq y$ and $y \succeq x$, which is written as $x \sim y$. If $x$ is weakly preferred over $y$ but we are not indifferent between the two then we also say that $x$ is strictly preferred over $y$, written as $x \succ y$.

We will see the type of preferences which can be expressed through filtering
and sorting is restricted. To characterize these restrictions, we first
introduce some properties. 

\begin{definition}[$\calA$-definable property] A subset of $X$ is said to be $\calA$-definable if it is a conjunction of atomic statements such as $a(x) \geq p$ and $a(x) \leq p$, where $a \in \calA$ and $p \in P$. 
\end{definition}

For instance, the set of alternatives which are cheaper than $\$10$ but have an average rating higher than $3$ is $\{ \price, \rating \}$-definable as $\price(x) \leq 10 \wedge \rating(x) \geq 3$.

\begin{definition}[$\calA$-definable ordering] An ordering on $X$ is said to be $\calA$-definable if it is the lexicographical combination of the orderings $\succeq_a$ and $\preceq_a$ for $a \in \calA$. 
\end{definition}

For instance, the ordering on $X$ which first orders elements by increasing $\price$ order and then by decreasing $\rating$ order (when two elements have the same price) is an $\calA$-definable ordering. When talking about $\calA$-definable ordering we will use the notation $\succeq_{(\alpha_1, a_1) \times \ldots \times (\alpha_n, a_n)}$ where $\alpha_i \in \{i, d\}$ and $a_i \in \calA$, denoting the ordering which first considers the $\alpha_1$-direction of property $a_1$, then the $\alpha_2$-direction of property $a_2$ and so on. In this way, the ordering above can be written concisely as $\succeq_{(i, \price) \times (d, \rating)}$. 

\begin{definition}[$\calA$-definable preference relation] A preference relation $A \subseteq X \times X$ is said to be $\calA$-definable if it can be expressed as $A(x,y) \equiv \neg F(y) \vee (F(x) \wedge x \succeq y)$ where $F$ is an $\calA$-definable property and $\succeq$ is an $\calA$-definable ordering.
\end{definition}

\begin{proposition} \label{prop-A-pref} For any $\calA$-definable property $F$ and any $\calA$-definable ordering $\succeq$ the relation
$$\neg F(y) \vee (F(x) \wedge x \succeq y)$$
is a preference relation.
\end{proposition} 
{\bf Proof}. Fix an $\calA$-definable filter $F$ and $\calA$-definable ordering $\succeq_{(\alpha_1, a_1) \times \ldots \times (\alpha_n, a_n)}$; and let 
\[ x \succeq y \; \equiv \; \neg F(y) \vee (F(x) \wedge x \succeq_{(\alpha_1, a_1) \times \ldots \times (\alpha_n, a_n)} y) \]
Let us first show that either $x \succeq y$ or $y \succeq x$. Indeed, for both of these to fail we should have $F(x)$ and $F(y)$ but not $x \succeq_{(\alpha_1, a_1) \times \ldots \times (\alpha_n, a_n)} y$ or $y \succeq_{(\alpha_1, a_1) \times \ldots \times (\alpha_n, a_n)} x$, which is impossible since the completeness of each individual ordering $\succeq_{(\alpha_i, a_i)}$ implies the completeness of the lexicographical ordering $\succeq_{(\alpha_1, a_1) \times \ldots \times (\alpha_n, a_n)}$. 
In order to show the transitivity of $\succeq$, assume $x \succeq y$ and $y \succeq z$. If $\neg F(z)$ we are done. Hence, suppose $F(z)$, and we must argue that $F(x)$ and $x \succeq_{(\alpha_1, a_1) \times \ldots \times (\alpha_n, a_n)} z$. Indeed, $F(z)$ and $y \succeq z$ gives us $F(y)$ and $y \succeq_{(\alpha_1, a_1) \times \ldots \times (\alpha_n, a_n)} z$, while $F(y)$ and $x \succeq y$ gives us $F(x)$ and $x \succeq_{(\alpha_1, a_1) \times \ldots \times (\alpha_n, a_n)} y$. These indeed imply $x \succeq_{(\alpha_1, a_1) \times \ldots \times (\alpha_n, a_n)} y$ by the transitivity of each individual ordering $\succeq_{(\alpha_i, a_i)}$. \hfill $\Box$

\subsection{Relating choice procedures and $\calA$-definable preference relations}

We now turn to the question how procedures and preferences are related. Let us
begin by introducing a candidate preference relation which can be derived from a
choice procedure. We will then show that this candidate is indeed an $\calA$-definable preference relation.

\begin{definition}[Preference relation derived from a procedure] Given a decision procedure $P \colon \List{X} \to \List{X}$ its \emph{preference relation} is defined as
\[ x \succeq_P y \;\; \equiv \;\; \exists l \in \List{X} (x,y \in l \wedge l' = P(l) \wedge \forall j (y = l'_j \to \exists i < j (x = l'_i))) \]
\end{definition}

In words, we say that $x$ is $P$-weakly preferred over $y$ if for some list
containing both $x$ and $y$ we have that the procedure $P$ either filters out
$y$, or it leaves both elements but ranks $x$ higher than $y$.

It is not obvious that $x \succeq_{P} y$ as defined above is a preference relation. Proposition
\ref{prop-A-pref} and the following theorem show that this is indeed the case.

\begin{theorem}[Soundness] \label{soundness} For any procedure $P \colon \List{X} \to \List{X}$ over the set of attributes $\calA$ there exists an $\calA$-definable preference relation $x \succeq y$ such that $x \succeq_P y \Leftrightarrow x \succeq y$.
\end{theorem}
{\bf Proof}. By the normal form theorem we have that $P$ is equivalent to a procedure of the form
\[ \sort{\alpha_1}{a_1} \circ \ldots \circ \sort{\alpha_n}{a_n} \circ \filter{\phi_1} \circ \ldots \circ \filter{\phi_m} \]
where $a_i$ are distinct attributes and $\alpha_i \in \{ i , d \}$ and $\phi_j$ are of the form $a \geq v$ or $a \leq v$. Let $F(x)$ be the $\calA$-definable property
\[ F(x) \equiv \bigwedge_{j=1}^m \phi_j \]
and let $\succeq_{(\alpha_1, a_1) \times \ldots \times (\alpha_n, a_n)}$ be the $\calA$-definable ordering. Finally, let $x \succeq y$ be the $\calA$-definable preference relation
\[ x \succeq y \equiv \neg F(y) \vee (F(x) \wedge x \succeq_{(\alpha_1, a_1) \times \ldots \times (\alpha_n, a_n)} y) \]
We claim that $x \succeq_P y \Leftrightarrow x \succeq y$. \\[1mm]
First, assuming $x \succeq_P y$ holds, i.e. there exists a list $l$ such that 
\[ x,y \in l \wedge l' = P(l) \wedge \forall j (y = l'_j \to \exists i < j (x = l'_i)). \]
If $\neg F(y)$ then $x \succeq y$ holds trivially and we are done. So let us assume $F(y)$. This means that $y$ passes all the filters of $P$ and hence $y \in l'$ and $x \in l'$, with $y$ coming later than $x$ on the resulting list. Since the sorting component of $P$ is $\sort{\alpha_1}{a_1} \circ \ldots \circ \sort{\alpha_n}{a_n}$, and $y$ comes later than $x$ in $l'$, it is clear that $x \succeq_{(\alpha_1, a_1) \times \ldots \times (\alpha_n, a_n)} y$. \\[1mm]
For the other direction, assume $\neg F(y) \vee (F(x) \wedge x \succeq_{(\alpha_1, a_1) \times \ldots \times (\alpha_n, a_n)} y)$. We must show that for some list $l$ containing $x$ and $y$ we have either
\[ y \not\in l' \quad\quad \mbox{or} \quad\quad y = l'_j \wedge x = l'_i \wedge i < j \]
where $l' = P(l)$. If $y$ does not pass one of the filters of $P$ this is an easy task. So let us assume $F(y)$. Our assumption then gives $F(x)$ and $x \succeq_{(\alpha_1, a_1) \times \ldots \times (\alpha_n, a_n)} y$. Hence, let $l$ be any list where $x$ is listed before $y$. Since we have $F(x)$ and $F(y)$, we obtain that $x, y \in l'$. Assume $x = l'_i$ and $y = l'_j$. Finally, since $x \succeq_{(\alpha_1, a_1) \times \ldots \times (\alpha_n, a_n)} y$, and the sorting operations of $P$ match this lexicographical ordering, we indeed have that $i < j$. \hfill $\Box$ \\

The converse of the above also holds: 

\begin{theorem}[Completeness] \label{completeness} For any $\calA$-definable preference relation $(X, \succeq)$ there is an $\calA$ procedure $P_{\succeq} \colon \List{X} \to \List{X}$ such that $x \succeq_{P_\succeq} y$ iff $x \succeq y$.
\end{theorem}
{\bf Proof}. Suppose the given $\calA$-definable preference relation is of the form
\[ x \succeq y \; \equiv \; F(y) \to (F(x) \wedge x \succeq_{(\alpha_1, a_1) \times \ldots \times (\alpha_n, a_n)} y) \]
for an $\calA$-definable property
\[ F(x) \equiv \bigwedge_{j=1}^m \phi_j \]
Define the procedure
\[ P_{\succeq} \equiv \sort{\alpha_1}{a_1} \circ \ldots \sort{\alpha_n}{a_n} \circ \filter{\phi_1} \circ \ldots \circ \filter{\phi_m} \]
We can show that $x \succeq_P y$ iff $x \succeq y$ using a similar argument to the proof of Theorem \ref{soundness}. First, assuming $x \succeq y$, i.e. 
\[ F(y) \to (F(x) \wedge x \succeq_{(\alpha_1, a_1) \times \ldots \times (\alpha_n, a_n)} y) \]
we need to produce a list $l$ containing both $x$ and $y$ such that 
\[ \forall j (y = l'_j \to \exists i < j (x = l'_i)) \]
where $l' = P_{\succeq}(l)$. Again, if $\neg F(y)$, meaning that $y$ does not pass the filter $F$, we can take $l$ to be any list containing $x$ and $y$, since $y$ will be filtered out and $y \not\in l'$. Assume, however, that $F(y)$, then by our assumption we must also have $F(x)$ and  $x \succeq_{(\alpha_1, a_1) \times \ldots \times (\alpha_n, a_n)} y$. But this indeed implies that in any list $l$ containing $x$ and $y$, with $x$ listed before $y$, we will have (for $l' = P(l)$) that $x, y \in l'$ and $x$ is be sorted ahead of $y$ in $l'$. The converse, that $x \succeq_{P_\succeq} y$ implies $x \succeq y$, also follows a similar pattern to the proof of Theorem \ref{soundness}. \hfill $\Box$ \\

To sum up: Theorems \ref{soundness} and \ref{completeness} show that there is  
 a one-to-one correspondence between choices arising from procedures and  $\calA$-definable preference relations.
For agents who have $\calA$-definable preferences, sorting and filtering is all they need. They can choose the first element and will get their optimal choice.  

\subsection{Complexity of choice procedure}

It is not only interesting that agents can use filtering and sorting
to choose optimally but doing so will be \emph{simple}.  
In order to be able to compare procedures based on sorting and filtering to other ways of optimizing
we need to define a benchmark case. 

\begin{definition}[Element-by-element optimization]
The agent compares the first two elements of the list and keeps the better
element. He proceeds by comparing the better element to the third element;
keeping the better element again. He continues in this way to the last element.   
\end{definition}

Consider a crude but for our purpose sufficient measure of simplicity: the number of steps needed to find the optimal alternative.
In the case of element-by-element optimization, or for that matter any procedure
which inspects the whole list, in order to find the optimal element in a list of
length $n$, the agent needs to make $n-1$
comparisons.

\begin{proposition}\label{prop:complexity}
Assume an agent holds $\calA$-definable preferences and the number of attributes
used by him is $N \leq | \cal A |$ and the length of the input list is $n$. If $N < (n-1)/3$ then choosing by filtering and
sorting is quicker than maximizing element by element.  
\end{proposition}
{\bf Proof}. The agent filters or sorts according to the attributes and then
chooses the first element on the resulting list. As the number of attributes is $N$, by Corollary \ref{cor-upper-bound},
the length of the procedure (in normal form) will be at most $3 N$. And if 
$3 N < n - 1$, choosing in this way will be quicker than the element-by-element optimization. \hfill $\Box$ \\

From the last result, it is clear that with increasing variety, the
decision support by filtering and sorting is more and more likely to quicken the
time needed to find the optimal element. 

So far we assumed agents hold specific
preferences \emph{and} the platform does provide the relevant attributes.
In the next section we turn to cases where one of these assumptions is violated
and study in which sense the available functionality still works approximately.

\section{Dealing with Missing Attributes}

Let us consider the case when the agent's preferences are defined on a strictly larger set of
attributes $\calA' \supset \calA$. For instance, suppose an online shop allows customers to sort
and filter with respect to $\price$ and $\rating$, but the customer also has some preferences over
the $\dimension$ (e.g. of a table) or the $\capacity$ (e.g. of an external hard disk). In this section, we discuss how
the agent might still be able to use the operations available to assist in their
choice.

\subsection{Fewer filters means more satisficing}

Consider first an agent who would like to choose the cheapest highest-rated hard disk with a capacity of at least 1TB, but
$\calA = \{ \price, \rating \}$. His preference can be fully captured by a procedure over the extended
set of attributes $\calA' = \{ \price, \rating, \capacity \}$, namely
\[ \sort{d}{\rating} \circ \sort{i}{\price} \circ \filter{\capacity \geq \textup{1TB}} \]
Over the smaller attribute set $\calA = \{ \price, \rating \}$, where the filtering operation $\filter{\capacity \geq \textup{1TB}}$ is missing, we can only approximate this procedure
as
\[ \sort{d}{\rating} \circ \sort{i}{\price} \]
What should the agent then do? He cannot filter the elements by $\capacity$. So,
he must manually inspect the list, mustn't he?

We claim: The agent can still use the available operations, namely
$\sort{d}{\rating} \circ \sort{i}{\price}$. Then, he can apply a satisficing
procedure on the set $\capacity \geq$ 1TB. In other words, the lack of a filtering
mechanism on a particular attribute can be replaced by a final
\emph{satisficing} step.

Satisficing can be formally defined as follows:

\begin{definition}[Satisficing] Consider an agent who has in mind a partition of the set of alternatives $X$, $\{S, X \setminus S \}$, with $S$ describing the set of alternatives which are satisfactory. A \emph{satisficing} $\List$-decision procedure consists of the agent scanning the list until a satisfactory element is found. If none is found the last element is chosen.
This procedure can be described by a simple recursive procedure as:
\begin{itemize}
   \item[] $\xi_S([x]) = x$ 
   \item[] $\xi_S(x:xs) =$ \mbox{if $x \in S$ then $x$ else $\xi_S(xs)$} 
\end{itemize}
\end{definition}

The lack of a filter operation can be replaced by a final satisficing step:

\begin{proposition} For any $\List$-decision procedure $P$ we have that 
$$(\first \circ P \circ \filter{a \geq v})(l) = (\xi_{\{x \colon a(x) \geq v \}} \circ P)(l)$$
for any list $l$ for which $(P \circ \filter{a \geq v})(l)$ is not empty.
\end{proposition}
{\bf Proof}. Fix $l \in \List{X}$ and assume $(P \circ \filter{a \geq v})(l)$ is not empty, so that $P(l)$ is not empty either. By Proposition \ref{prop-swap} $(P \circ \filter{a \geq v})(l) = (\filter{a \geq v} \circ P)(l)$. Hence, $(\first \circ P \circ \filter{a \geq v})(l)$ is the first element of $P(l)$ which satisfies $a \geq v$. That is exactly what the satisficing procedure $\xi_{\{x \colon a(x) \geq v \}}$ would do to the list $P(l)$. \hfill $\Box$ \\

According to the above proposition, if the agent follows a satisficing procedure and can use the 
existing attributes on the platform, then filtering will produce a list of
results which are all satisfactory to the agent. He then can choose any of the
resulting list and will be satisfied.
In other words, if an attribute is missing, the agent can use a
satisficing procedure instead. Hence, after filtering with the available
attributes, he can then use a satisficing procedure to find an element that
satisfies the missing attribute.

In passing, we note the following relationship between satisficing, filtering, and maximization:

\begin{proposition}
If the satisficing set $S$ consists of the maximal elements with respect to an $\calA$-definable preference relation $\succeq$, then $(\first \circ P_{\succeq})(l) = \xi_S(l)$, where $P_{\succeq}$ is defined in the proof of Proposition \ref{completeness}, and assuming $l$ is such that $l_i \in S$ for some index $i$.
\end{proposition}
{\bf Proof}. Suppose the given $\calA$-definable preference relation is of the form
\[ x \succeq y \; \equiv \; F(y) \to (F(x) \wedge x \succeq_{(\alpha_1, a_1) \times \ldots \times (\alpha_n, a_n)} y) \]
for an $\calA$-definable property $F(x) \equiv \bigwedge_{j=1}^m \phi_j$ so that $P_{\succeq}$ is the procedure
\[ P_{\succeq} \equiv \sort{\alpha_1}{a_1} \circ \ldots \sort{\alpha_n}{a_n} \circ \filter{\phi_1} \circ \ldots \circ \filter{\phi_m} \]
Let $l$ be a list such that $l_i \in S$ for some index $i$. It is clear that the first element $l'_0$ of $l' = P_{\succeq}(l)$ will be such that $F(l_0')$ and $l_0'$ will be maximal in $l'$ with respect to the order $(\alpha_1, a_1) \times \ldots \times (\alpha_n, a_n)$. It remains to see that for any other element in $l$ which also satisfies $F$ and is maximal with respect to $(\alpha_1, a_1) \times \ldots \times (\alpha_n, a_n)$, has to come at position $l_k'$ for some $k > 0$ by the assumption ($S_3$) on the sorting operation. \hfill $\Box$ \\

That is, if an agent is happy with any of the alternatives in $S$, and if $S$ are the optimal elements according to an $\calA$-definable preference relation, then the agent can replace the satisficing procedure by the much quicker procedure of applying $P_{\succeq}$ and then selecting the first element. 

Turning back to the filter, sort, satisficing procedure, why should an agent
consider a satisficing step at all? Because often it is quicker than
element-by-element maximization!

\begin{proposition} Suppose $\calA \subset \calA'$. If the agent holds $\calA'$-definable preferences (using $N$ attributes) and $n > n' + 3 N$, where $n$ is the length of the list of alternatives, $n'$ is the length of the list after filtering, using the filter and sort procedure available in $\calA$, followed by satisficing on the attribute
not listed is quicker than element-by-element maximization. 
\end{proposition}
{\bf Proof}. The agent filters or sorts according to the attributes in $\calA$ and then
satisfices w.r.t. the missing attribute $\calA' \backslash \calA$. By Corollary \ref{cor-upper-bound}, each attribute is used at most three times.
Hence, if the filter and sort combination sufficiently reduce the number
of alternatives, here measured by the difference between the initial list $n$ and
the resulting list $n'$, then satisficing will be at least as quick as
maximization of the reduced list. In all but the case where there is only one
element that is satisfactory and happens to be the last element on the list,
satisficing will be quicker than element-by-element maximization. \hfill $\Box$ \\

\subsection{Less sorting means more local maximization}

Let us now turn to the case where some sorting operations are missing.  Consider an agent would like to choose the cheapest disk with the highest capacity. This again can be described by a procedure over the attribute set $\calA' = \{ \price, \rating, \capacity \}$ as
\[ \sort{d}{\capacity} \circ \sort{i}{\price} \]
but over the smaller set $\calA = \{ \rating, \capacity \}$ all we can do is sort by $\capacity$. How should the agent then find the cheapest
disk with the highest capacity? It might seem at first that the agent would need to scan the whole list
in order to find such disk. But since only the disk with highest capacity should be considered, the search for the
cheapest disk is done locally only within the products with highest capacity!

Therefore, the lack of some sorting mechanism implies the need for the agent to scan
some portion of the results performing a \emph{local} maximization procedure.

\begin{definition}[Local maximization] Consider an agent who would like to maximize with respect to some attribute $a$, but will only consider the initial elements that have the same $a_1, \ldots, a_n$-values for some other attributes $a_1, \ldots, a_n$. A \emph{local maximization} $\List$-decision procedure consists of the agent scanning the list until a different $a_1, \ldots, a_n$-values is found, and then taking the $a$-maximal element seen up to that point. This procedure can also be described by a simple recursive algorithm as:
\begin{itemize}
   \item[] $\nu'_{a;a_1\ldots a_n}(b)([\,]) = b$ 
   \item[] $\nu'_{a;a_1\ldots a_n}(b)(x:xs) =$ \mbox{if $\vec{a}(x) = \vec{a}(b)$ then $\nu_S(\max_a\{x,b\})(xs)$ else $b$} 
\end{itemize}
and $\nu_{a;a_1 \ldots a_n}(x:xs) = \nu'_{a; a_1\ldots a_n}(x)(xs)$, where $\vec{a}(x) = \vec{a}(b)$ means $a_i(x) = a_i(b)$ for all $1 \leq i \leq n$.
\end{definition}

The following proposition is immediate:

\begin{proposition} For any attributes $a$ and $a'$ and procedure $P$, we have that
$$(\first \circ \sort{\alpha_1}{a_1} \circ \ldots \circ \sort{\alpha_n}{a_n} \circ \sort{d}{a} \circ P)(l) = (\nu_{a; a_1\ldots a_n} \circ \sort{\alpha_1}{a_1} \circ \ldots \circ \sort{\alpha_n}{a_n} \circ P)(l)$$
i.e. performing a local maximization over $a$ on an initial segment of a list which has the same $a_1 \ldots a_n$ values, is equivalent to sorting the list in decreasing $a$-order first, then by $a_1 \ldots a_n$ order, and taking the first element. 
\end{proposition}

Why should an agent do the local maximization? Because often it is the quickest way to
find the best alternative! 

As long as the number of alternatives is large, and preferences can be
defined on attributes  -- even if some of them are not listed, the filtering and
sorting help to choose optimally. And if the
number of alternatives increases, they help even more to quickly find the best
alternative. 

\subsection{Approximating through filtering and sorting}

What can an agent do, if his preferences cannot be directly expressed through the available functionality? He will need to scan the whole list of
alternatives, if he wants to fully optimize. Depending on the number of alternatives, this can be a daunting task.

In such a situation, the agent may deviate from full optimization and revert to
an approximation. Obviously, there are many ways he could shortcut the problem
(and many have been discussed in the literature). There is no way to generally
state what an approximation should be.

Note though, one prominent alternative to full optimization is the
satisficing procedure introduced above. An agent following this procedure inspects the items in
the list until he finds an object which is satisfactory for him. And as noted
above, satisficing can be implemented very fast using filtering.

\section{More General Procedures}

So far, we have explored the relationship between a limited online
environment, the functionality offered for a customer, and what kind of preference
relations can be directly expressed through this functionality. In the last
section, we considered the case where the environment is not expressive enough.

Consider a property buyer is looking for the cheapest house which \emph{either} has three bedrooms \emph{or} is located within 1 mile from the train station. So both types of houses are substitutes for him, and he would like to find the cheapest of these. 
Assuming the attributes $\calA = \{ \bedrooms, \price, \distance \}$ are
available online, the desired query would need to involve a \emph{union} of two
filtering operations. But our available composition of filters only captures the
\emph{intersection} operation, i.e. the operation
\[ \first \circ \sort{i}{\price} \circ \filter{\bedrooms \geq 3} \circ \filter{\distance \leq 1} \]
finds the cheapest house, which has 3 or more bedrooms \emph{and} is within one mile from the train station.

At first, it might seem, the agent can simply perform two queries: one looking at houses with three bedrooms, and another looking at
houses near the station. But with two lists one can no longer easily apply further operations, such as
sorting the results according to price and year of construction, etc.

Note this issue would be resolved by the platform, if it offered this particular
combination as a single attribute $\threebedor$. Although artificial in the particular example, it might make sense to extend the attribute set with this compound attribute if it is 
commonly requested. More generally, the set of attributes $\calA$ available online does not need to contain just atomic or even simple composite attributes but could also include much richer attributes.
In this section, however, we go in a different direction. We extend the platform
with a general union operation and consider the implications.

\subsection{General filters and decision procedures}

\begin{definition}[General decision filter] Define the intersection and union of filter operations as
\begin{itemize}
	\item $\filter{\phi_1} \cap \filter{\phi_2} \colon \List{X} \to \List{X}$ as the filter operation which will discard the elements of the list unless it satisfies both filter conditions $\phi_1$ and $\phi_2$.
	\item $\filter{\phi_1} \cup \filter{\phi_2} \colon \List{X} \to \List{X}$ as the filter operation which will discard the elements of the list unless it satisfies at least one of the filter conditions $\phi_1$ or $\phi_2$.
\end{itemize}
For each attribute $a \colon X \to V$ let us call the filter operations $\filter{a \geq v}$ and $\filter{a \leq v}$ atomic filters. We then call a \emph{general decision filter} any combination of the atomic filters by means of unions and intersections.
\end{definition}

For instance, we can filter out the houses that have a price tag of at most  \euro $2000$ and, have 3 or more bedrooms \emph{or} are within one mile from the train station, as
\[ \filter{\price \leq 2000} \cap (\filter{\bedrooms \geq 3} \cup \filter{\distance \leq 1}) \]

\begin{proposition} Any general decision filter is equivalent to one which is the intersection of the union of atomic filters.
\end{proposition}
{\bf Proof}. A basic result of propositional logic says that any propositional formula is logically equivalent to one in \emph{conjunctive normal form}, i.e. it is the disjunction of conjunctions. When applied to the above, we see that it is sufficient to consider general decision filters which are obtained via the intersection of unions, and that this is enough to capture any general decision filter. \hfill $\Box$ \\

It is therefore sufficient to extend an online platform with an operation for the union of atomic filters, since the intersection is already captured by the sequential composition of procedures. 

\begin{definition}[General decision procedure] Given a set of attributes $\calA$, the general decision procedures over $\calA$ are those obtained by compositing sorting operations with a general decision filter. 
\end{definition}

Since our normal form theorem shows that we can, without loss of generality,
assume that filtering operations are done before sorting, these general decision
procedures indeed generalizes our previous notion of a decision procedure.

We begin with a more general notion of $\calA$-definable property,

\begin{definition}[General $\calA$-definable property] A subset of $X$ is said to be general $\calA$-definable if it can be expressed in propositional logic from the atomic predicates $a(x) \geq v$ and $a(x) \leq v$, where $a \colon X \to V$ is an attribute in $\calA$ and $v \in V$. 
\end{definition}

as well as a more general notion of $\calA$-definable preference relation.

\begin{definition}[General $\calA$-definable preference relation] A preference relation $A \subseteq X \times X$ is said to be general $\calA$-definable if it can be expressed as $A(x,y) \equiv \neg F(y) \vee (F(x) \wedge x \succeq y)$ where $F$ is a general $\calA$-definable property and $\succeq$ is an $\calA$-definable ordering.
\end{definition}

\subsection{Generalized filters and definable preferences}

Mirroring our work in Section \ref{sec:modelling_pref}, we now relate the
generalized version of choice procedures to the richer preference structures we
introduced before. Again, we split the results into two theorems. To begin with, any general
procedure defines a preference relation:

\begin{theorem}[General soundness] \label{general soundness} For any \emph{general} procedure $P \colon \List{X} \to \List{X}$ over the set of attributes $\calA$ there exists a \emph{general} $\calA$-definable preference relation $x \succeq y$ such that $x \succeq y \Leftrightarrow x \succeq_P y$.
\end{theorem}
{\bf Proof}. A general produce $P \colon \List{X} \to \List{X}$ is of the form
\[ \sort{\alpha_1}{a_1} \circ \ldots \circ \sort{\alpha_n}{a_n} \circ (\Sigma_1 \cap \ldots \cap \Sigma_m) \]
where $\Sigma_i = \filter{\phi_{i,1}} \cup \ldots \cup \filter{\phi_{i,m_i}}$. Let $F(x)$ be the general $\calA$-definable property
\[ F(x) \equiv \bigwedge_{i=1}^m \bigvee_{k=1}^{m_i} \phi_{i,k} \]
and let $\succeq_{(\alpha_1, a_1) \times \ldots \times (\alpha_n, a_n)}$ be the $\calA$-definable ordering. Finally, let $x \succeq y$ be the general $\calA$-definable preference relation
\[ x \succeq y \equiv \neg F(y) \vee (F(x) \wedge x \succeq_{(\alpha_1, a_1) \times \ldots \times (\alpha_n, a_n)} y) \]
As in Theorem \ref{soundness}, we can show that $x \succeq y \Leftrightarrow x \succeq_P y$. \\[1mm]
First, assuming $x \succeq_P y$ holds, i.e. there exists a list $l$ such that 
\[ x,y \in l \wedge l' = P(l) \wedge (y \in l' \to (y = l'_j \wedge x = l'_i \wedge i < j)). \]
If $\neg F(y)$ then $x \succeq y$ holds trivially and we are done. So let us assume $F(y)$. This means that $y$ passes all the filters of $P$ and hence $y \in l'$ and $x \in l'$, with $y$ coming later than $x$ on the resulting list. Since the sorting component of $P$ is $\sort{\alpha_1}{a_1} \circ \ldots \circ \sort{\alpha_n}{a_n}$, and $y$ comes later than $x$ in $l'$, it is clear that $x \succeq_{(\alpha_1, a_1) \times \ldots \times (\alpha_n, a_n)} y$. \\[1mm]
For the other direction, assume $\neg F(y) \vee (F(x) \wedge x \succeq_{(\alpha_1, a_1) \times \ldots \times (\alpha_n, a_n)} y)$. We must show that for some list $l$ containing $x$ and $y$ we have either
\[ y \not\in l' \quad\quad \mbox{or} \quad\quad y = l'_j \wedge x = l'_i \wedge i < j \]
where $l' = P(l)$. If $y$ does not pass one of the filters of $P$ this is an easy task. So let us assume $F(y)$. Our assumption then gives $F(x)$ and $x \succeq_{(\alpha_1, a_1) \times \ldots \times (\alpha_n, a_n)} y$. Hence, let $l$ be any list where $x$ is listed before $y$. Since we have $F(x)$ and $F(y)$, we obtain that $x, y \in l'$. Assume $x = l'_i$ and $y = l'_j$. Finally, since $x \succeq_{(\alpha_1, a_1) \times \ldots \times (\alpha_n, a_n)} y$, and the sorting operations of $P$ match this lexicographical ordering, we indeed have that $i < j$. \hfill $\Box$

The converse also holds: Given any general $\calA$-definable preference
relation, we can find a procedure which implements the optimal choice according
to the preferences.

\begin{theorem}[General completeness] \label{general completeness} For any \emph{general} $\calA$-definable preference relation $(X, \succeq)$ there is a \emph{general} $\calA$-procedure $P \colon \List{X} \to \List{X}$ such that $x \succeq y$ iff $x \succeq_P y$.
\end{theorem}
{\bf Proof}. Again, the proof is similar to that of Theorem \ref{completeness}. Suppose the given general $\calA$-definable preference relation is of the form
\[ x \succeq y \; \equiv \; F(y) \to (F(x) \wedge x \succeq_{(\alpha_1, a_1) \times \ldots \times (\alpha_n, a_n)} y) \]
for a general $\calA$-definable property
\[ F(x) \equiv \bigwedge_{i=1}^m \bigvee_{k=1}^{m_i} \phi_{i,k} \]
Define the procedure
\[ P_{\succeq} \equiv \sort{\alpha_1}{a_1} \circ \ldots \sort{\alpha_n}{a_n} \circ (\Sigma_1 \cap \ldots \cap \Sigma_m) \]
where $\Sigma_i = \filter{\phi_{i,1}} \cup \ldots \cup \filter{\phi_{i,m_i}}$. It is easy to verify that $x \succeq y$ iff $x \succeq_P y$ (cf. proof of Theorem \ref{completeness}). \hfill $\Box$ \\

Theorems \ref{general soundness} and \ref{general completeness} show that the
one-to-one relationship between choice procedures and preferences we derived
before can be extended to the more general case. And, as before, the
availability of these support functions speeds up the choice process.

\section{Conclusion}

In this paper we show that online decision support system can be used to choose
rationally. The offered functionality speeds up the decision process. In Simon's terms, the
choice environment contains complexity so that the individual's procedure
operating on that environment can be simpler.

In the case of simple filter and sorting functionality this only works if the
type of preferences is restricted. We also show how the operations can be
extended so that more general notions of preference relations are captured by
filter and sorting functionality. This line of thought can be extended
further: For each set of operations we can ask what type of preferences can be
expressed. But we can also ask the opposite question: What kind of operations do
we need to effectively express preferences of a specific structure?

If one considers current technological developments, it becomes evident that
support systems become more expressive (or at least that is the design goal of
many companies). In the light of our results, this will lead to individual procedures being simpler
while being able to choose according to complex preferences. As noted
before, the better such systems the more variety can be handled without imposing
more complexity on the consumer's choice process.

\bibliographystyle{aea}
\bibliography{complexity}

@article{masatlioglu2013choice,
  title={Choice by iterative search},
  author={Masatlioglu, Yusufcan and Nakajima, Daisuke},
  journal={Theoretical Economics},
  volume={8},
  number={3},
  pages={701--728},
  year={2013},
  publisher={Wiley Online Library}
}

@article{caplin2011search,
  title={Search, choice, and revealed preference},
  author={Caplin, Andrew and Dean, Mark},
  journal={Theoretical Economics},
  volume={6},
  number={1},
  pages={19--48},
  year={2011},
  publisher={Wiley Online Library}
}

@article{OlivaZahn2018,
   author = {Oliva, Paulo and Zahn, Philipp},
    title = "{The choice environment, constraints, and rational procedures}",
  journal = {ArXiv e-prints},
archivePrefix = "arXiv",
   eprint = {1801.03483},
 primaryClass = "cs.GT",
     year = 2018,
    month = jan,
   adsurl = {http://adsabs.harvard.edu/abs/2018arXiv180103483O}
}

@article{Todd2003,
title = "Bounding rationality to the world",
journal = "Journal of Economic Psychology",
volume = "24",
number = "2",
pages = "143 - 165",
year = "2003",
note = "The Economic Psychology of Herbert A. Simon",
issn = "0167-4870",
doi = "https://doi.org/10.1016/S0167-4870(02)00200-3",
url = "http://www.sciencedirect.com/science/article/pii/S0167487002002003",
author = "Peter M Todd and Gerd Gigerenzer"
}

@article{Rubinstein1998definable,
title = "Definable preferences: An example",
journal = "European Economic Review ",
volume = "42",
number = "3–5",
pages = "553 - 560",
year = "1998",
note = "",
issn = "0014-2921",
doi = "http://dx.doi.org/10.1016/S0014-2921(98)00008-7",
url = "http://www.sciencedirect.com/science/article/pii/S0014292198000087",
author = "Ariel Rubinstein"
}

@article{Simon1955,
author = {Simon, Herbert A.}, 
title = {A Behavioral Model of Rational Choice},
volume = {69}, 
number = {1}, 
pages = {99-118}, 
year = {1955}, 
doi = {10.2307/1884852}, 
URL = {http://qje.oxfordjournals.org/content/69/1/99.abstract}, 
eprint = {http://qje.oxfordjournals.org/content/69/1/99.full.pdf+html}, 
journal = {The Quarterly Journal of Economics} 
}

@article{simon1956,
  title={Rational choice and the structure of the environment.},
  author={Simon, Herbert A},
  journal={Psychological review},
  volume={63},
  number={2},
  pages={129},
  year={1956},
  publisher={American Psychological Association}
}

@article{Rubinstein2006b,
author = {Rubinstein, Ariel and Salant, Yuval},
isbn = {15557561},
issn = {15557561},
journal = {Theoretical Economics},
pages = {3--17},
title = {{A model of choice from lists}},
url = {http://tspace.library.utoronto.ca/handle/1807/4780},
volume = {1},
year = {2006}
}

@article{Salant2008,
author = {Salant, Yuval and Rubinstein, Ariel},
journal = {Review of Economic Studies},
pages = {1287--1296},
title = {{(A,f) : Choice with Frames}},
volume = {75},
year = {2008}
}

@article{iyengar2000choice,
  title={When choice is demotivating: Can one desire too much of a good thing?},
  author={Iyengar, Sheena S and Lepper, Mark R},
  journal={Journal of personality and social psychology},
  volume={79},
  number={6},
  pages={995},
  year={2000},
  publisher={American Psychological Association}
}

@article{de2008offering,
  title={Offering online recommendations with minimum customer input through conjoint-based decision aids},
  author={De Bruyn, Arnaud and Liechty, John C and Huizingh, Eelko KRE and Lilien, Gary L},
  journal={Marketing science},
  volume={27},
  number={3},
  pages={443--460},
  year={2008},
  publisher={INFORMS}
}

@article{liu2018semantic,
  title={A semantic approach for estimating consumer content preferences from online search queries},
  author={Liu, Jia and Toubia, Olivier},
  journal={Marketing Science},
  volume={37},
  number={6},
  pages={930--952},
  year={2018},
  publisher={INFORMS}
}

@article{gilbride2004choice,
  title={A choice model with conjunctive, disjunctive, and compensatory screening rules},
  author={Gilbride, Timothy J and Allenby, Greg M},
  journal={Marketing Science},
  volume={23},
  number={3},
  pages={391--406},
  year={2004},
  publisher={INFORMS}
}

@article{haubl2000consumer,
  title={Consumer decision making in online shopping environments: The effects of interactive decision aids},
  author={H{\"a}ubl, Gerald and Trifts, Valerie},
  journal={Marketing science},
  volume={19},
  number={1},
  pages={4--21},
  year={2000},
  publisher={INFORMS}
}

@article{bronnenberg2016zooming,
  title={Zooming in on choice: How do consumers search for cameras online?},
  author={Bronnenberg, Bart J and Kim, Jun B and Mela, Carl F},
  journal={Marketing science},
  volume={35},
  number={5},
  pages={693--712},
  year={2016},
  publisher={INFORMS}
}

@article{chen2017sequential,
  title={Sequential search with refinement: Model and application with click-stream data},
  author={Chen, Yuxin and Yao, Song},
  journal={Management Science},
  volume={63},
  number={12},
  pages={4345--4365},
  year={2017},
  publisher={INFORMS}
}

@article{koulayev2014search,
  title={Search for differentiated products: identification and estimation},
  author={Koulayev, Sergei},
  journal={The RAND Journal of Economics},
  volume={45},
  number={3},
  pages={553--575},
  year={2014},
  publisher={Wiley Online Library}
}

@article{kim2017probit,
  title={The probit choice model under sequential search with an application to online retailing},
  author={Kim, Jun B and Albuquerque, Paulo and Bronnenberg, Bart J},
  journal={Management Science},
  volume={63},
  number={11},
  pages={3911--3929},
  year={2017},
  publisher={INFORMS}
}

@article{hortaccsu2004product,
  title={Product differentiation, search costs, and competition in the mutual fund industry: A case study of S\&P 500 index funds},
  author={Horta{\c{c}}su, Ali and Syverson, Chad},
  journal={The Quarterly journal of economics},
  volume={119},
  number={2},
  pages={403--456},
  year={2004},
  publisher={MIT Press}
}

@article{de2012testing,
  title={Testing models of consumer search using data on web browsing and purchasing behavior},
  author={De los Santos, Babur and Horta{\c{c}}su, Ali and Wildenbeest, Matthijs R},
  journal={American economic review},
  volume={102},
  number={6},
  pages={2955--80},
  year={2012}
}

@article{santos2017search,
  title={Search with learning for differentiated products: Evidence from e-commerce},
  author={Santos, Babur De Los and Horta{\c{c}}su, Ali and Wildenbeest, Matthijs R},
  journal={Journal of Business \& Economic Statistics},
  volume={35},
  number={4},
  pages={626--641},
  year={2017},
  publisher={Taylor \& Francis}
}

@article{moraga2013semi,
  title={Semi-nonparametric estimation of consumer search costs},
  author={Moraga-Gonz{\'a}lez, Jos{\'e} Luis and S{\'a}ndor, Zsolt and Wildenbeest, Matthijs R},
  journal={Journal of Applied Econometrics},
  volume={28},
  number={7},
  pages={1205--1223},
  year={2013},
  publisher={Wiley Online Library}
}

@article{shi2021path,
  title={The Path to Click: Are You on It?},
  author={Shi, Savannah Wei and Trusov, Michael},
  journal={Marketing Science},
  volume={40},
  number={2},
  pages={344--365},
  year={2021},
  publisher={INFORMS}
}

@article{hauser2010disjunctions,
  title={Disjunctions of conjunctions, cognitive simplicity, and consideration sets},
  author={Hauser, John R and Toubia, Olivier and Evgeniou, Theodoros and Befurt, Rene and Dzyabura, Daria},
  journal={Journal of Marketing Research},
  volume={47},
  number={3},
  pages={485--496},
  year={2010},
  publisher={SAGE Publications Sage CA: Los Angeles, CA}
}

\end{document}